\documentclass[10pt,final]{iopart}
\usepackage{iopams}  
\usepackage{graphicx}
\begin{document}


\title[Multiresolution techniques for the detection of gravitational-wave
       bursts]
      {Multiresolution techniques for the detection of gravitational-wave
       bursts}

\author{S Chatterji, L Blackburn, G Martin and E Katsavounidis}

\address{MIT LIGO Laboratory, NW17-161, Cambridge, MA 02139, USA}

\eads{\mailto{shourov@ligo.mit.edu} and \mailto{lindy@ligo.mit.edu}}


\begin{abstract}
We present two search algorithms that implement logarithmic tiling of the
time-frequency plane in order to efficiently detect astrophysically unmodeled
bursts of gravitational radiation.  The first is a straightforward application
of the dyadic wavelet transform.  The second is a modification of the windowed
Fourier transform which tiles the time-frequency plane for a specific $Q$.  In
addition, we also demonstrate adaptive whitening by linear prediction, which
greatly simplifies our statistical analysis.
This is a methodology paper that aims to describe the techniques for
identifying significant events as well as the necessary pre-processing
that is required in order to improve their performance. For this reason
we use simulated LIGO noise in order to illustrate the methods and to
present their preliminary performance.
\end{abstract}


\submitto{\CQG}



\section{Introduction}
\label{sec:introduction}

Gravitational wave bursts\cite{ref:thorne1} are short (less than one second)
transients of gravitational radiation with poorly known waveforms.  They are one
of the distinct types of signals LIGO\cite{ref:gonzalez} and other
interferometric and resonant mass detectors\cite{ref:thorne2} are pursuing.

A search for such signals is generally a time-frequency search, since time and
frequency information of the candidate events are essential for a multi-detector
coincidence analysis as well as for relating events to astrophysical sources.
One of the well established tools for such an analysis is the short time Fourier
transform (also known as the windowed Fourier transform) developed by Denis
Gabor in 1946. This approach applies standard Fourier transforms to short
windowed segments of the original time series to create a uniform time-frequency
map with a time resolution and frequency resolution that depends upon the
window's duration.  However, it is impossible to pick a window duration
appropriate for all frequencies.  Low frequency signals require long duration
windows to accumulate sufficient frequency information, while high frequency
signals are better isolated in time using short duration windows.  Time and
frequency localization of transient signals is therefore not easily achieved
with the short time Fourier transform.  Here we have also assumed that burst
waveforms are limited in $Q$; that is, that the number of oscillations fall
within a particular range regardless of frequency.

In this introductory methods paper we explore multiresolution techniques for
the detection of transients that rely on the dyadic wavelet
transform\cite{ref:daubechies,ref:mallat} and the constant $Q$
transform\cite{ref:stockwell,ref:brown}.  Both methods search for a
statistically significant excess of signal power in the time-frequency plane.
However, contrary to existing excess power
searches\cite{ref:abbott,ref:anderson,ref:sylvestre,ref:klimenko}, we implement
a bank of filters that produce a logarithmic tiling of the time-frequency plane.
This provides increasing time resolution at higher frequencies and thus
naturally addresses the time-frequency localization problem.

This paper outlines these new methods and is organized as follows.
Section~\ref{sec:bursts} defines a representation independent parameterization
of gravitational-wave bursts and defines an optimal signal to noise ratio, which
motivates our use of a multiresolution approach.  Section~\ref{sec:wavelet}
presents the basics of the continuous and discrete wavelet transforms, while
Section~\ref{sec:q} presents the continuous and discrete $Q$ transform as a
modification of the short time Fourier transform.  Section~\ref{sec:lpef}
demonstrates how linear prediction can be used to whiten background noise prior
to transform analysis.  Section~\ref{sec:events} describes methods for the
selection of significant events from the output of the proposed algorithms, and
Section~\ref{sec:efficiency} presents preliminary detection efficiencies of our
search algorithms applied to Gaussian and sine-Gaussian bursts injected into
simulated LIGO detector noise.  Tuning and application of the proposed
algorithms to search for a broader set of signals in data from the first LIGO
science runs is currently in progress.


\section{Gravitational-Wave Bursts}
\label{sec:bursts}

To further motivate the development of multiresolution techniques, we first
consider the parameterization and detectability of gravitational-wave bursts
that are well localized in both time and frequency. The more general case of
non-localized bursts can be treated as a superposition of localized bursts.


\subsection{The Parameterization of Gravitational-Wave Bursts}
\label{sec:bursts:parameterization}

An arbitrary gravitational wave burst may be represented in both the time-domain
and the frequency-domain by the Fourier transform pair $h(t)$ and
$\tilde{h}(f)$.  For bursts that are square-integrable, we may also define a
representation independent characteristic squared amplitude,
\begin{equation}
\label{eqn:characteristic_amplitude}
\| h \|^2 =
\int_{-\infty}^{+\infty} | h(t) |^2 \, dt =
\int_{-\infty}^{+\infty} | \tilde{h}(f) |^2 \, df.
\end{equation}
For bursts that are well localized in both time and frequency, it is then
meaningful to define a central time, central frequency, duration, and
bandwidth:
\begin{eqnarray}
t_c & = \int_{-\infty}^{+\infty} t \frac{| h(t) |^2}{\| h \|^2} \, dt
\qquad \qquad
\sigma_t^2 & = \int_{-\infty}^{+\infty} (t - t_c)^2 \frac{| h(t) |^2}{\| h \|^2}  \, dt \\
f_c & = 2 \int_{0}^{\infty} f \frac{| \tilde{h}(f) |^2}{\| h \|^2} \, df
\qquad \qquad
\sigma_f^2 & = 2 \int_{0}^{\infty} (f - f_c)^2 \frac{| \tilde{h}(f) |^2}{\| h \|^2} \, df.
\end{eqnarray}
It can be shown that the duration and bandwidth, when defined in this way, obey
an uncertainty relation of the form,
\begin{equation}
\label{eqn:uncertainty}
\sigma_t \sigma_f \geq \frac{1}{4 \pi},
\end{equation}
which we will also take as an approximate criterion for bursts that are well
localized in the time-frequency plane.
Finally, we define the dimensionless quality factor of a burst,
\begin{equation}
\label{eqn:q}
Q = f_c / \sigma_f,
\end{equation}
which for well localized bursts, is simply a measure of the burst's aspect
ratio in the time-frequency plane.


\subsection{The Detectability of Gravitational-Wave Bursts}
\label{sec:bursts:detectability}

In general, a search algorithm for gravitational-wave bursts projects the data
under test onto a basis constructed to span the space of plausible bursts.  The
optimal measurement occurs when a member of this basis exactly matches a
gravitational-wave burst.  In this case we achieve a signal to noise ratio,
$\rho$, which for narrowband bursts or a flat detector noise spectrum, is simply
the ratio of the total energy content of the signal, $\| h \|^2$, to the in-band
one-sided power spectral density, $S_h(f)$, of the detector noise:
\begin{equation}
\label{eqn:snr}
\rho^2 = \int_{0}^{\infty} \frac{2 | \tilde{h}(f) |^2}{S_h(f)} \, df
\simeq \frac{\| h \|^2}{S_h(f)}.
\end{equation}
Thus, $\| h \|$, which has units of strain per square root Hz, is a convenient
quantity for evaluating the detectability of narrowband bursts since it is
directly comparable to the amplitude spectral density of detector noise.

It is important to note, however, that the above signal to noise ratio is only
achieved when a member of the measurement basis closely matches the signal in
the time-frequency plane.  Otherwise, the measurement encompasses either too
little signal or too much noise, resulting in a loss in the measured signal to
noise ratio.  Since bursts are naturally characterized by their $Q$, we
therefore choose to tile the time frequency plane by selecting a measurement
basis which directly targets bursts within a finite range of $Q$. This naturally
leads to a logarithmic tiling of the time frequency plane in which individual
measurement pixels are well localized signals, all with the same $Q$.


\section{The Wavelet Transform}
\label{sec:wavelet}


\subsection{The Continuous Wavelet Transform}
\label{sec:wavelet:continuous}

In the preceding section, we motivate a search for bursts which projects the
data under test onto a basis of well localized bursts of constant $Q$.  The
Wavelet transform\cite{ref:mallat} by construction satisfies these requirements, and
is defined by the integral,
\begin{equation}
\label{eqn:cwt}
W_{f}(u,s) =
\int_{-\infty}^{+\infty} f(t) \,
\frac{1}{\sqrt{s}} \, \psi^{\star}
\left( \frac{t-u}{s} \right) \, dt,
\end{equation}
where the wavelet, $\psi$, is a time-localized function of zero average.

The coefficients $W_{f}(u,s)$ are evaluated continuously over times, $u$, and
scales, $s$.  Our ability to resolve in time and frequency is then determined
by the properties $\psi$ assumes at each scale.  At large scale, $\psi$ is
highly dilated yielding improved frequency resolution at the expense of time
resolution.  At small scale, we achieve good time resolution with large
uncertainty in frequency.  However, since the number of oscillations is fixed,
the resulting $Q$ is constant over all scales.


\subsection{The Discrete Dyadic Wavelet Transform}
\label{sec:wavelet:discrete}

For the case of discrete data, a computationally efficient algorithm exists
for calculating wavelet coefficients over scales that vary as powers of two: $
s \in \{ 2^{j-1} \, | \, j \in \mathbb{Z}^{+} \}$.
This is the dyadic wavelet transform, which can be implemented for a limited
family of wavelets using conjugate mirror filters.  The filters consist of a
high pass filter, $\hat{H}$, and low pass filter, $\hat{L}$, which can be
applied in a cascade to obtain the wavelet coefficients.  Beginning with the
original time series, $A_0$, of length $N$, two sequences of length $N/2$ are
obtained by application of the high pass and low pass filters followed by
down-sampling.  The sequence of detail coefficients, $D_j$, and approximation
coefficients, $A_j$, are defined at each level, $j$, of the decomposition by
\begin{equation}
\label{eqn:dwt}
D_{j} = \hat{H}(A_{j-1})
\hfill
\mbox{and}
\hfill
A_{j} = \hat{L}(A_{j-1}).
\hspace{1.0in}
\end{equation}
The detail coefficients for scale $s$, where $s = 2^{j-1}$, calculated in this
manner are the same as the wavelet coefficients obtained from
Equation~\ref{eqn:cwt}.  If $N$ is a power of two, so that $N = 2^m$, the final
approximation sequence will be $A_m$.

The simplest dyadic wavelet is the Haar function:
\begin{equation}
\label{eqn:haar}
\psi^{\mbox{\scriptsize Haar}}(t) = \left\{ \begin{array}{rl}
  1 & 0 \leq t < 1/2 \\
 -1 & 1/2 \leq t < 1 \\
  0 & \mbox{otherwise.}
\end{array} \right.
\end{equation}
The corresponding high pass and low pass filters are
\begin{equation}
\hat{H}^{\mbox{\scriptsize Haar}} = [+1, -1] / \sqrt{2}
\hfill
\mbox{and}
\hfill
\hat{L}^{\mbox{\scriptsize Haar}} = [+1, +1] / \sqrt{2},
\hspace{0.5in}
\end{equation}
from which we see that the detail coefficients are related to the differences of
each pair of points in the parent series, while the approximation coefficients
are related to the averages of each pair.

A technique called wavelet packets can be used to further decompose the detail
coefficients into smaller frequency subbands of higher $Q$. The {\em Waveburst}\cite{ref:klimenko}
search method currently implements a full wavelet packet
decomposition which results in a uniform time-frequency plane with time-frequency
cell sizes equivalent to those of the highest scale in the dyadic decomposition.
It is possible, however, to choose a different subset of the wavelet packet
coefficients which approximate a more constant $Q$ time-frequency plane by
further decomposing the detail coefficents only a constant number of times at each
scale. This would give much more flexibility to detect high $Q$ signals which do 
not match well to the mother wavelet $\psi$.


\section{The $Q$ Transform}
\label{sec:q}

The $Q$ transform is a modification of the standard short time Fourier
transform in which the analysis window duration varies inversely with
frequency.  It is similar in design to the continuous wavelet transform and
also permits efficient computation for the case of discrete data.  However,
since reconstruction of the data sequence is not a concern, we permit
violation of the zero mean requirement.  In addition, the discrete $Q$
transform is not constrained to frequencies which are related by powers of
two.


\subsection{The Continuous Q Transform}
\label{sec:q:continuous}

We begin by projecting the sequence under test, $x(t)$, onto windowed complex
sinusoids of frequency $f$:
\begin{equation}
\label{eqn:qtransform}
x(\tau, f) = \int_{-\infty}^{+\infty} x(t) w(t - \tau, f)
e^{-i 2 \pi f t} \, dt.
\end{equation}
Here $w(t - \tau, f)$ is a time-domain window centered on time $\tau$ with a
duration that is inversely proportional to the frequency under
consideration\cite{ref:stockwell,ref:brown}.
The $Q$ transform may also be expressed in an alternative
form\cite{ref:stockwell}, which allows for efficient computation.
\begin{equation}
\label{eqn:qtransform:alt}
x(\tau, f) = \int_{-\infty}^{+\infty} \tilde{x}(\phi + f) \tilde{w}^{*}(\phi, f)
e^{+i 2 \pi \phi \tau} \, d\phi.
\end{equation}
Thus, the $Q$ transform at a specific frequency is obtained by a standard
Fourier transform of the original time series, a shift in frequency,
multiplication by the appropriate frequency domain window function, and an
inverse Fourier transform.  The benefit of Equation~\ref{eqn:qtransform:alt}
is that the Fourier transform of the original time series need only be
computed once.  We then perform the inverse Fourier transform only for the
logarithmically spaced frequencies that we are interested in.


\subsection{The Discrete Q Transform}
\label{sec:q:discrete}

We may also adapt the $Q$ transform to the case of discrete data.  In this case,
Equation~\ref{eqn:qtransform} takes the form:
\begin{equation}
\label{eqn:dqt}
x[m, k] = \sum_{n = 0}^{N - 1} x[n] w[n - m, k] e^{- i 2 \pi n k / N}.
\end{equation}
As was the case for the continuous $Q$ transform, we may express the discrete
$Q$ transform in the alternative form:
\begin{equation}
\label{eqn:dqt:alt}
x[m, k] = \sum_{l = 0}^{N - 1} \tilde{x}[l + k] \tilde{w}^{*}[l, k]
e^{+ i 2 \pi m l / N}.
\end{equation}
This allows us to take advantage of the computational efficiency of the fast
Fourier transform for both the initial transform as well as the subsequent
inverse transforms for each desired value of $k$.

Due to the implementation of the discrete $Q$ transform in
Equation~\ref{eqn:dqt:alt}, the window function is most conveniently defined in
the frequency domain.  In particular, we choose a frequency domain Hanning
window, which has finite support in the frequency domain while still providing
good time-frequency localization.


\section{Adaptive whitening by linear prediction}
\label{sec:lpef}

In this section, we present linear
prediction\cite{ref:makhoul,ref:haykin,ref:cuoco,ref:searle} as a tool for
whitening gravitational-wave data prior to analysis by both the dyadic wavelet
transform and the discrete $Q$ transform.  For both transforms, the statistical
distribution of transform coefficients is well known for the case of input white
noise.  Whitened data therefore permits a single predefined test for
significance, which greatly simplifies the subsequent identification of
candidate bursts.  We also note that linear predictive whitening is a necessary
component of current cross-correlation\cite{ref:cadonati,ref:mohanty} and
time-domain\cite{ref:mcnabb} searches for gravitational-wave bursts in LIGO
data.

Linear prediction assumes that the $n^{\mbox{\scriptsize th}}$ sample of a
sequence is well modeled by a linear combination of the previous $M$ samples.
We thus define the predicted sequence,
\begin{equation}
\label{eqn:lpef:prediction}
\hat{x}[n] = \sum_{m=1}^{M} c[m] x[n-m],
\end{equation}
and the corresponding prediction error sequence,
\begin{equation}
\label{eqn:lpef:error}
e[n] = x[n] - \hat{x}[n].
\end{equation}
The coefficients $c[m]$ are then chosen to minimize the mean squared prediction
error,
\begin{equation}
\label{eqn:lpef:sigmae}
\sigma_e^2 = \frac{1}{N} \sum_{n = 1}^{N} | e[n] |^2,
\end{equation}
over a representative training sequence of length $N$.
Assuming that $x[n]$ is a stationary stochastic process, this results in
the well known Yule-Walker\cite{ref:makhoul} equations,
\begin{equation}
\label{eqn:lpef:yulewalker}
\sum_{k = 1}^{M} r[m - k] c[m] = r[k]
\qquad
\mbox{for}
\qquad
1 \leq k \leq M,
\end{equation}
where $r[k]$ is the autocorrelation of the training sequence $x[n]$ evaluated at
lag $k$.  In practice, we choose the biased autocorrelation estimate,
\begin{equation}
\label{eqn:lpef:autocorrelation}
r[k] = \frac{1}{N} \sum_{n = |k|}^{N} x[n] x[n - |k|],
\end{equation}
which guarantees the existence of a solution to the Yule-Walker equations.

When this process is applied to gravitational-wave data, the resulting
prediction error sequence is composed primarily of sample to sample uncorrelated
white noise.  However, it also contains any unpredictable transient signals
which were present in the original data sequence.  Thus, the prediction error
sequence is the whitened data sequence which we desire to subject to further
analysis by the algorithms proposed in this paper.  We therefore define the
linear predictor error filter as the $M^{\mbox{\scriptsize th}}$ order finite
impulse response filter, which when applied to a data sequence, $x[n]$, returns
the corresponding prediction error sequence, $e[n]$:
\begin{equation}
\label{eqn:lpef:application}
e[n] = \sum_{m=0}^{M} b[m] x[n-m].
\end{equation}

The choice of the filter order, $M$, is then determined by the measurement
resolution of subsequent analysis.  For a given filter order, the resulting
prediction error sequence will be uncorrelated on time scales shorter than $M$
samples.  In the frequency domain, this corresponds to a spectrum which is white
on frequency scales greater than a characteristic bandwidth,
\begin{equation}
\label{eqn:lpef:bandwidth}
\Delta f \gtrsim \frac{f_{s}}{M},
\end{equation}
where $f_s$ is the sample frequency of the data.  Thus, we can whiten to any
desired frequency resolution by selecting the appropriate filter order, $M$.  In
practice, we simply choose the minimum filter order which is sufficient for the
subsequent analysis.  For the training length, $N$, we choose to train over
times which are long compared to the typical gravitational-wave bursts we are
searching for.

\begin{figure}
\begin{center}
\scalebox{1.0}[0.8]{\includegraphics[width=2.5in]{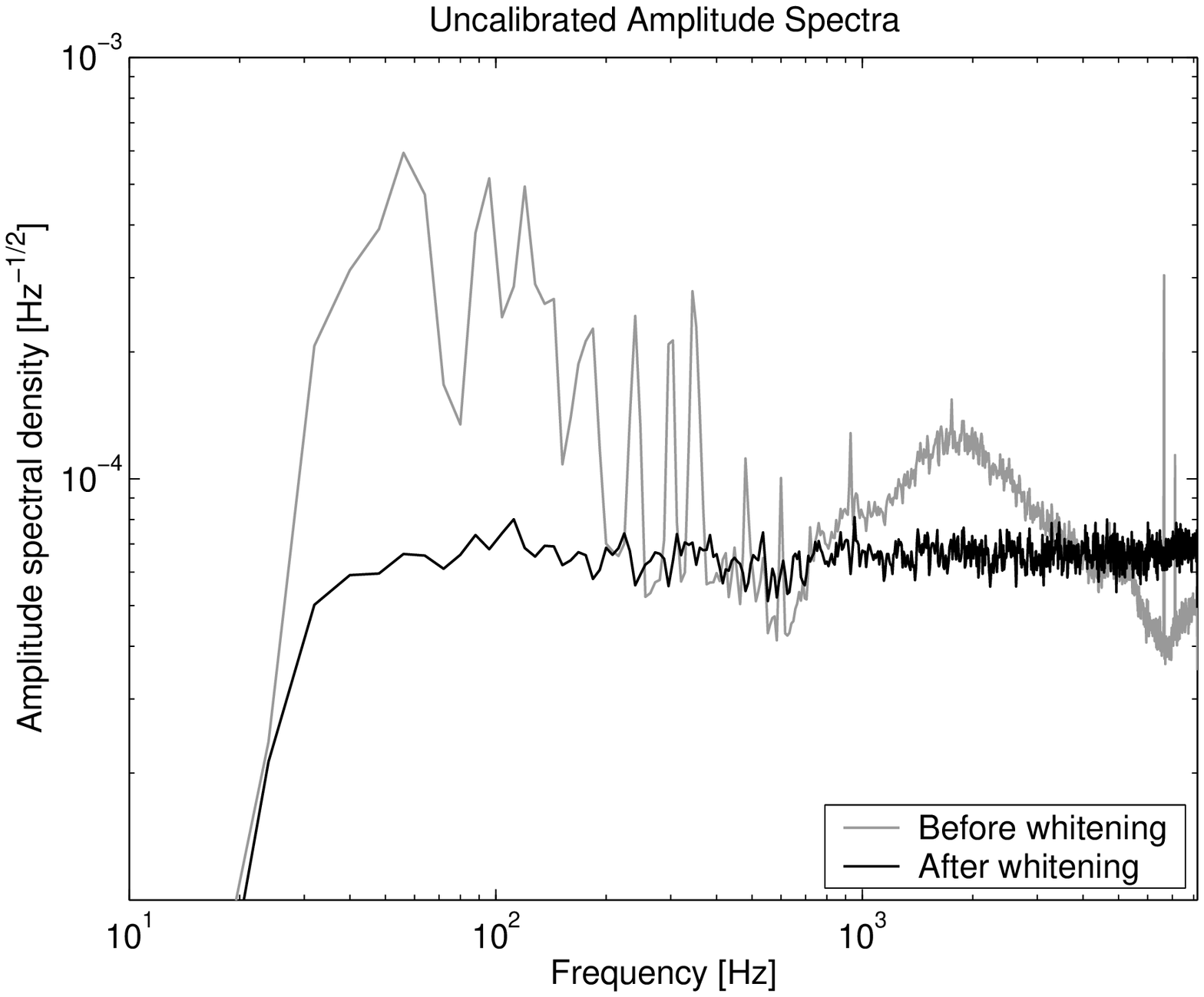}}
\scalebox{1.0}[0.8]{\includegraphics[width=2.5in]{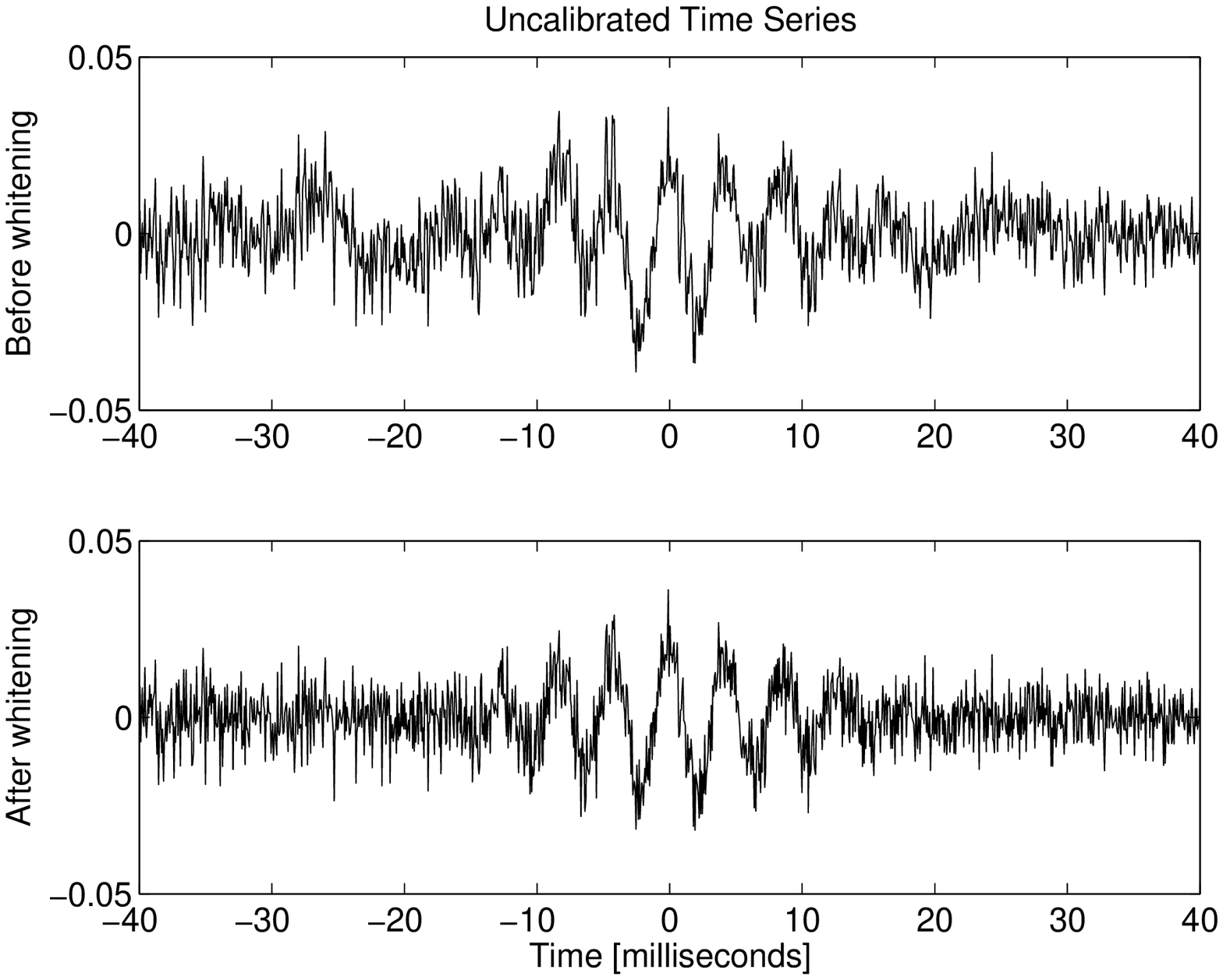}}
\end{center}
\caption{Example application of zero-phase linear predictive whitening to 48
seconds of sample uncalibrated gravitational-wave data from the second LIGO
science run.  In this case, the data was first high pass filtered at 64 Hz.  A
linear prediction error filter with 4 Hz whitening resolution was then trained
on the first 16 seconds of data and applied to the subsequent 32 seconds of
data.  The ability of this process to whiten the data is evident in the left
hand plot which shows the amplitude spectra of the data before and after
whitening, when measured with an 8 Hz resolution in accordance with
Equation~\ref{eqn:lpef:bandwidth}.  In addition, the right hand plot shows the
effect of the same linear prediction error filter when applied to a simulated
gravitational-wave burst injected into the data.  In this case, we inject a 235
Hz sine-Gaussian with a $Q$ of 12.7 and signal to noise ratio of 20 as described
in Section~\ref{sec:efficiency}.  Note the absence of a time delay between the
whitened and unwhitened burst.}
\label{fig:lpef}
\end{figure}

In order to avoid timing errors due to the unspecified phase response of linear
prediction error filters, we construct a new filter from the unnormalized
autocorrelation of the original filter coefficients:
\begin{equation}
\label{eqn:lpef:zerophase}
b^{\prime}[k] = \sum_{m = |k|}^{M} b[m] b[m - |k|].
\end{equation}
While this new filter has the desired property of zero phase
response\cite{ref:oppenheim}, its magnitude response has been squared.  In
practice, the construction of the zero phase filter is carried out in the
frequency domain, where it is also possible to take the square root of the
resulting frequency domain filter before transforming back to the time domain.
An example application of zero-phase linear prediction to sample
gravitational-wave data and a simulated gravitational-wave burst is shown in
Figure~\ref{fig:lpef}.


\section{Event Selection}
\label{sec:events}

We now seek to identify statistically significant events in the resulting dyadic
wavelet transform and discrete $Q$ transform coefficients.


\subsection{Dyadic Wavelet Transform Event Selection}
\label{sec:events:wavelet}

\begin{figure}
\begin{center}
\scalebox{1.0}[0.8]{\includegraphics[width=2.5in]{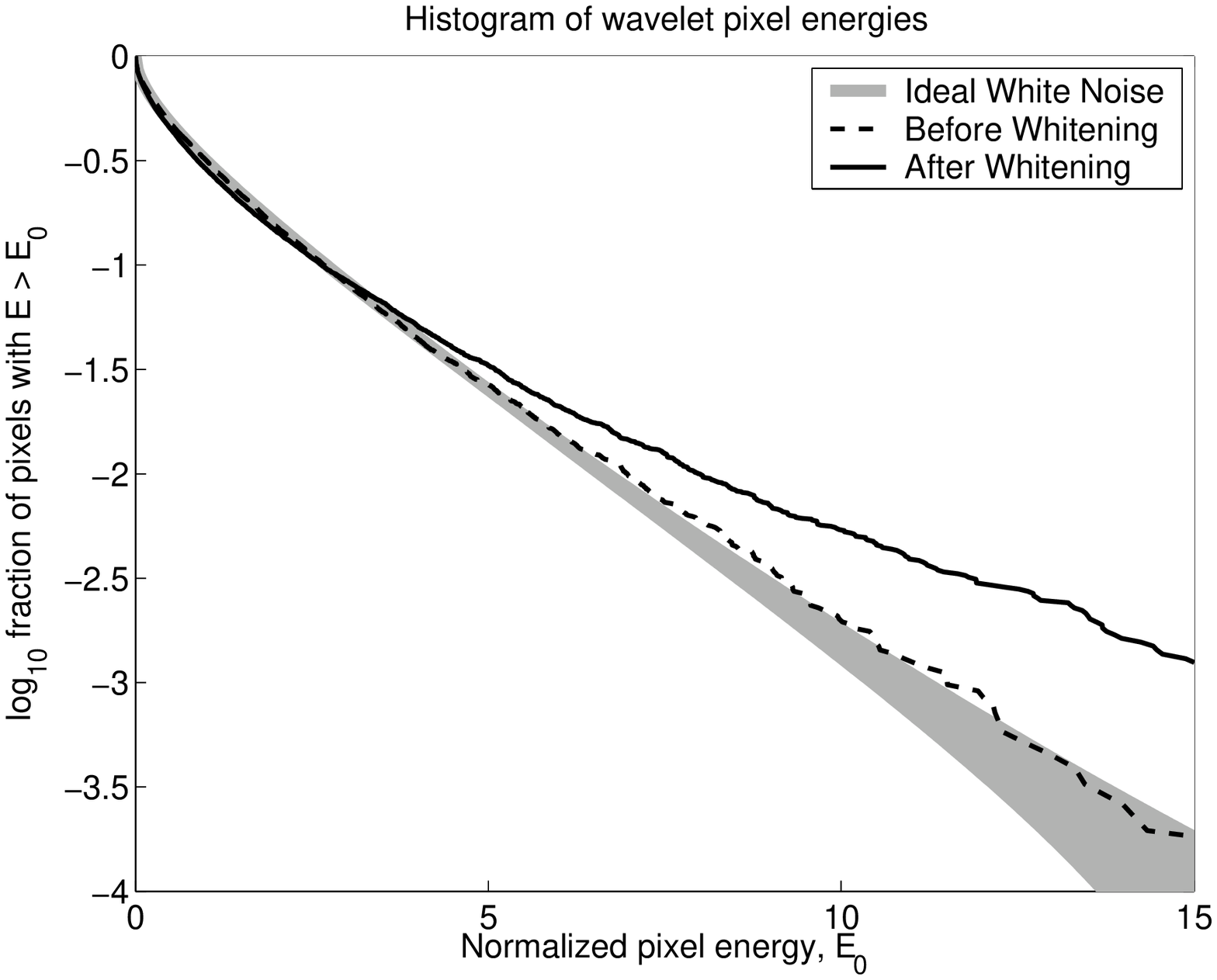}}
\scalebox{1.0}[0.8]{\includegraphics[width=2.5in]{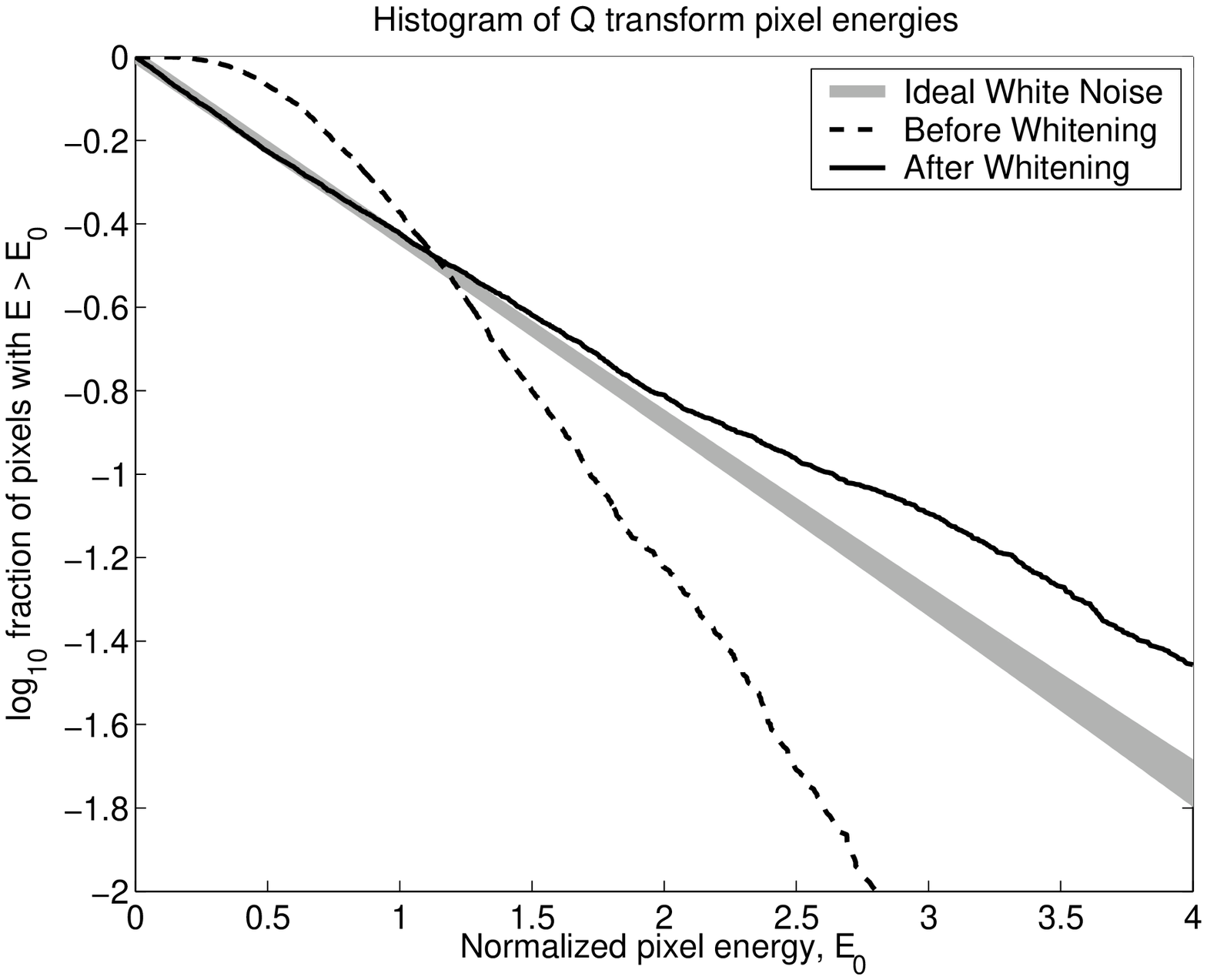}}
\end{center}
\caption{The distribution of dyadic wavelet transform (left) and discrete $Q$
transform (right) pixel energies in the scale and frequency bin encompassing 600
Hz for the data of Figure~\ref{fig:lpef}.  The distributions before (dashed) and
after (solid) linear predictive whitening are compared to the theoretically
expected distributions for ideal white noise (shaded).  The whitened data show
good agreement with theory except at high pixel energy, where the unpredictable
non-stationary behavior (glitches) of the gravitational-wave data becomes
apparent.  For the discrete $Q$ transform, the inconsistency of the unwhitened
data with theoretical white noise is due to the coherent 600 Hz power line
harmonic, which results in a Rician rather than exponential distribution of
pixel energies.  For the discrete Haar wavelet transform, the presence of a
coherent signal results in an apparent, but false, white noise distribution.
This is due to the use of a real valued wavelet which is sensitive to the
relative phase between the signal and wavelet.  This effect actually masks true
non-stationarities, which are only exposed after linear predictive whitening
removes the coherent signal content.}
\label{fig:histogram}
\end{figure}

We assume that linear predictive whitening (Section~\ref{sec:lpef}) has been
applied to the data prior to the dyadic wavelet transform.  Then, according to
the central limit theorem, for a sufficiently large scale, $j$, the wavelet
coefficients, $D_j$, within the scale will approach a zero-mean Gaussian
distribution with standard deviation $\sigma_j$.  We therefore define the
sequence of squared normalized coefficients, or normalized pixel energies at
scale $2^j$,
\begin{equation}
\label{eqn:energy:dwt}
E_j = D^2_j / \sigma^2_j,
\end{equation}
which is chi-squared distributed with one degree of freedom
(Figure~\ref{fig:histogram}).

It is then simply a matter of thresholding on the energy of individual pixels,
$\epsilon_{ij} \in E_j$, to identify statistical outliers.  We may also choose
to cluster nearby pixels to better detect bursts which deviate from the dyadic
wavelet tiling of the time-frequency plane.  For a randomly selected cluster $C$
of $N$ pixels, we define the total cluster energy,
\begin{equation}
\label{eqn:cluster:dwt}
E_C = \sum_{(i,j) \in C}{\epsilon_{ij}},
\end{equation}
which is also chi-square distributed, but with $N$ degrees of freedom.  It is
therefore possible to select a threshold energy to achieve a desired white noise
false event rate.  Alternatively, sharp transients may also be detected by
simply searching for vertical clusters of pixels within the top $P_j$ percent of
pixel energies in each scale.

\subsection{Discrete $Q$ Transform Event Selection}
\label{sec:events:dqt}

For the discrete $Q$ transform, we also make use of the central limit theorem by
assuming that the data has first been whitened using linear prediction.  For
pixels of sufficiently long duration, the pixel energies,
\begin{equation}
\label{eqn:energy:dqt}
\epsilon[m,k] = | x[m,k] |^2,
\end{equation}
at a particular frequency, $k$, will then approach an exponential distribution
(Figure~\ref{fig:histogram}).  Thus for each pixel, we have a measure of
significance:
\begin{equation}
P(\epsilon > \epsilon[m,k]) =
\exp \left( -\epsilon[m,k] / \mu_k \right),
\end{equation}
where $\mu_k$ is the mean pixel energy at frequency $k$.  As with the discrete
wavelet transform, we may also consider clusters of $N$ pixels and select a
threshold energy in order to achieve a desired white noise false event rate.
However in this case, the cluster energy is chi-square distributed with $2N$
degrees of freedom.

To efficiently detect bursts over a range of $Q$, we perform multiple $Q$
transforms, each of which produces a time-frequency plane tiled for a particular
$Q$ within our range of interest.
To best estimate the parameters of candidate
bursts, we then select the most significant non-overlapping pixels among all $Q$
planes using a simple exclusion algorithm.
Pixels are considered in decreasing
order of significance and any pixel which overlaps in time or frequency with a
more significant pixel is discarded.  Thus, for each localized candidate burst,
only the single pixel which best represents the burst parameters is reported.
This does not exclude the possibility of clustering pixels from non-localized
bursts, since the pixels which pass both the initial significance threshold as
well as the exclusion algorithm will represent the strong localized features of
such bursts. Reducing a non-localized burst to its strong, localized features
has the additional benefit of a much tighter time-frequency coincidence criterion
when multiple detectors are used.


\section{Detection efficiencies over simulated LIGO noise}
\label{sec:efficiency}

Finally, we evaluate the detection efficiencies of the proposed algorithms over
simulated LIGO noise.  The simulated data are composed of line sources and
Gaussian distributed white noise passed through shaping filters to produce a
power spectrum which matches the typical H1 detector noise spectrum during the
second LIGO science run\cite{ref:s2spectra}.  Assuming optimal orientation, we
evaluate the detection efficiency for each waveform at various $\| h \|$ by
injecting 256 such bursts at random times into the simulated data (see Figure
3).  The simulated data are first filtered by a 6th order 64 Hz Butterworth high
pass filter followed by a linear prediction error filter with whitening
resolution of 16 Hz.  However, we do not model the non-stationary behavior of
real LIGO noise, which largely determines the false event rate of existing
search algorithms.  The algorithms considered in this paper are equally
applicable to the identification of statistically significant transient events
in auxiliary interferometer or environmental data.  So, they may also prove
useful for identifying potential vetoes.

\begin{figure}
\begin{center}
\includegraphics[width=2.5in]{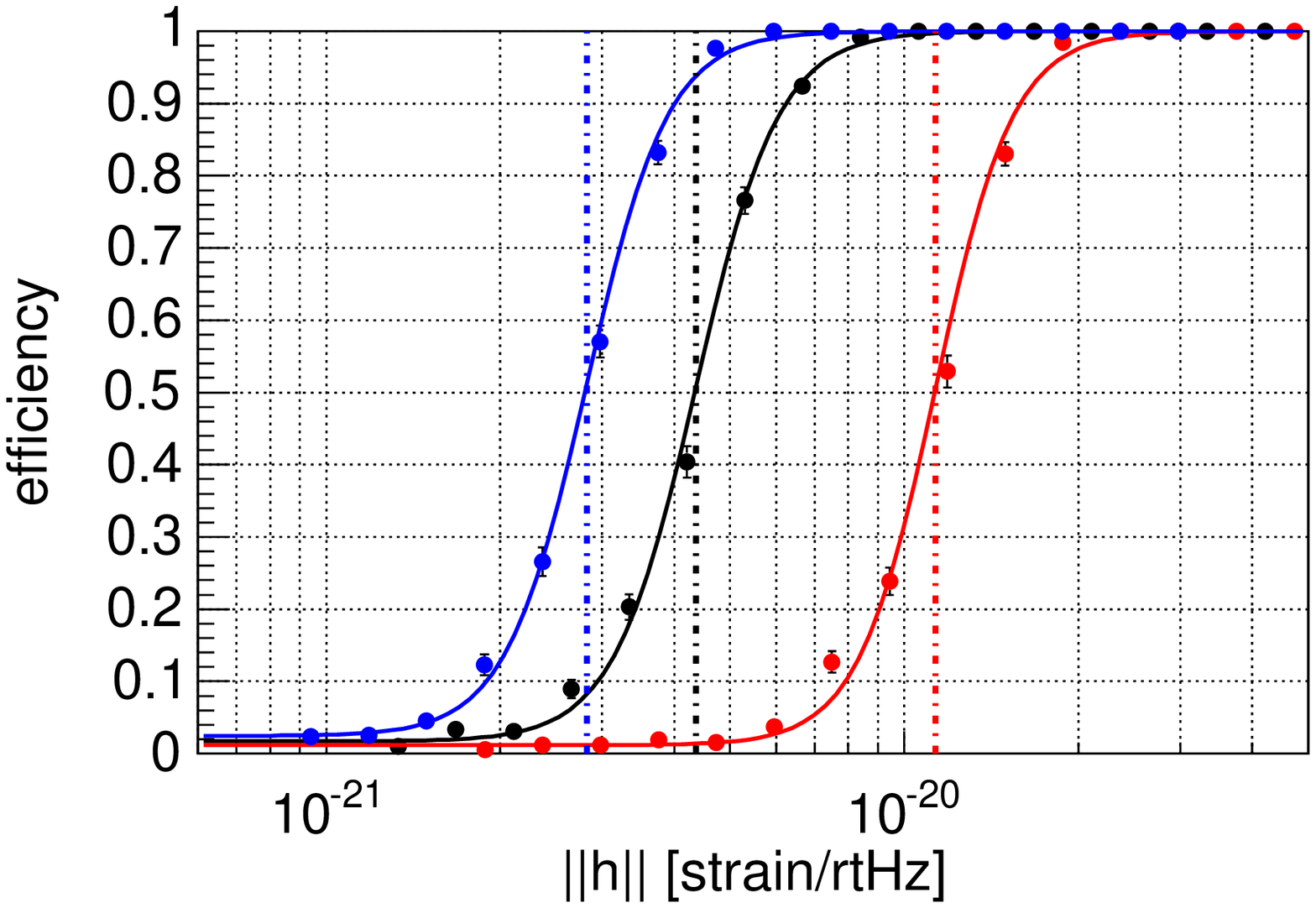}
\includegraphics[width=2.5in]{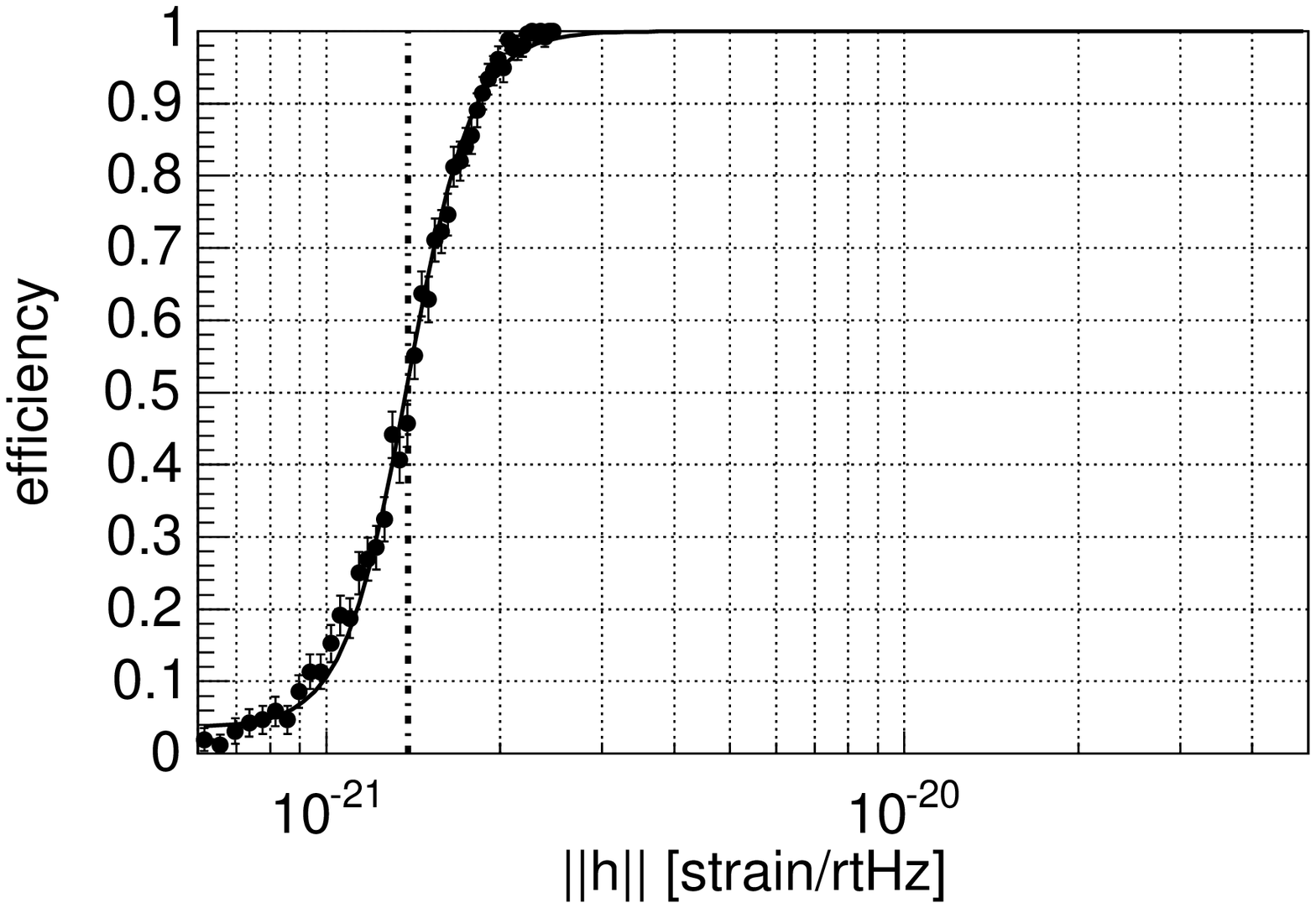}
\end{center}
\caption{Detection efficiency versus injected characteristic strain amplitude
for the dyadic wavelet transform (left) and the discrete $Q$ transform (right).
The detection efficiency of the dyadic wavelet transform is shown for three
durations of Gaussian bursts with $\sigma_t$'s of
0.35, 0.71, and 1.41 milliseconds as shown from left to right.  The detection
efficiency of the $Q$ transform is shown for a sine-Gaussian burst
with a $Q$ of 12.7 and central frequency,
$f_c$, of 275 Hz.}
\label{fig:efficiency}
\end{figure}

The low $Q$ Haar wavelet is best suited to broadband signals, and for a
representative broadband waveform, we use Gaussian
injections of the form $h(t) = h_0 \exp (- t^2 / 4 \sigma_t^2)$, for
$\sigma_t$'s of 0.35, 0.71, and 1.41 ms.  The filtered data are passed through
the discrete wavelet transform and event selection algorithm as described in
Section~\ref{sec:events:wavelet} for a false event rate of 0.5 Hz.  We allow
clusters to be composed of a single pixel, or a combination of two pixels
adjacent or overlapping in time and adjacent in scale for scales, $j$, of 5, 6,
and 7.  For $\sigma_t$'s of 0.35, 0.71, and 1.41 ms, we obtain 50\% efficiencies
for $\| h \|$ equal to $2.8\times10^{-21}$, $4.4\times10^{-21}$, and
$1.1\times10^{-20}$ Hz$^{-1/2}$, which correspond to signal to noise ratios,
$\rho$, of 3.6, 3.7, and 3.5.

Since the discrete $Q$ transform targets well localized bursts, we demonstrate
its efficiency for sine-Gaussian bursts of the form, $h(t) = h_0 \exp ( - t^2 /
4 \sigma_t^2 ) \sin ( 2 \pi f_c t )$, with a $Q$ of 12.7 and central frequency,
$f_c$, of 275 Hz.  The discrete $Q$ transform is applied to the whitened data
with approximately 75 percent pixel overlap to search for single pixel bursts
within the frequency band 64--4096 Hz and with $Q$'s in the vicinity of 10.6,
14.1, and 17.7.  Candidate events are then identified as described in
Section~\ref{sec:events:dqt} in order to obtain a 1 Hz white noise false rate.
The resulting 50\% detection efficiency occurs at an $\| h \|$ of $1.5\times
10^{-21}$ Hz$^{-1/2}$, corresponding to a $\rho$ of 3.0.


\section{Conclusions}

We have described two transient-finding algorithms using logarithmic
tiling of the time-frequency plane and examined their applicability
to the detection of gravitational wave bursts. The methods work more
efficiently when data are whitened and for this we have introduced a
data conditioning algorithm based on linear prediction.
As a first step in quantifying their performance, we have used
simulated noise data according to the LIGO noise spectra from its
second science run. Moreover, {\it{ad hoc}} waveform morphologies
were used to simulate the presence of a burst signal over the noise.
This allowed us to measure the algorithms' efficiency  at a fixed
false alarm rate of O(1Hz). The results are encouraging and we are
continuing the computation for mapping the efficiency of the algorithms
as a function of their false alarm rate.

Needless to say, even if the shaped LIGO noise with some line features
that we used for our simulation is a step forward, it does remain
unrealistic. Real interferometric data have richer line structure,
non-stationary and non-gaussian character that may be far from our
assumptions; at the same time though, it is extremely hard to simulate.
All of these make the perfomance estimates that we made using our
simulation model optimistic.
The use though of mathematically well described noise allows the
straightforward reproduction and comparison of our results with other
methods. An end-to-end pipeline using the aforementioned methods for
whitening and transient-finding on data collected by the LIGO instruments
is currently in development. This will provide the ultimate test of
the performance of the algorithms and will be reported separately
within the context of the LIGO Scientific Collaboration (LSC) data
analysis ongoing research.

\vspace{0.1in}
\noindent
This work was supported by the US National Science Foundation under Cooperative
Agreement No. PHY-0107417.



\end{document}